\documentclass[11pt]{article}
\usepackage{cite}
\usepackage{framed}
\usepackage{tikz,graphicx,graphics,color}
\usepackage{amsmath,amssymb,amsfonts}
\newcommand{\rem}[1]{}
\providecommand\bnabla{\boldsymbol{\nabla}}

\newsavebox{\astrutbox}
\sbox{\astrutbox}{\rule[-5pt]{0pt}{20pt}}

\newcommand{\bwt}{\begin{widetext}}
\newcommand{\ewt}{\end{widetext}}

\DeclareMathAlphabet{\mathbi}{OML}{cmm}{b}{it} 
\newcommand{\non}{\nonumber}

\newtheorem{lemma}{Lemma}

\newcommand{\tildel}{\tilde{\delta}_{m}}
\newcommand{\bx}{\mathbi{x}}

\newcommand{\bel}{\begin{equation}\label}
\newcommand{\ee}{\end{equation}}
\newcommand{\beq}{\begin{eqnarray}\label} 
\newcommand{\eeq}{\end{eqnarray}} 
\newcommand{\bc}{\begin{center}} 
\newcommand{\ec}{\end{center}} 
\newcommand{\ben}{\begin{enumerate}}
\newcommand{\een}{\end{enumerate}}
\newcommand{\bit}{\begin{itemize}}
\newcommand{\eit}{\end{itemize}}
\newcommand{\I}{\int_{\mathcal{V}}}
\newcommand{\bom}{\boldsymbol{\omega}}

\newcommand{\brot}{\bom_{rot}}

\newcommand{\bk}{\boldsymbol{\hat{k}}}
\newcommand{\bu}{\boldsymbol{u}}

\newcommand{\shalf}{{\ensuremath{\scriptstyle\frac{1}{2}}}}
\newcommand{\quart}{{\ensuremath{\scriptstyle\frac{1}{4}}}}

\newcommand{\onethird}{{\ensuremath{\scriptstyle\frac{1}{3}}}}
\newcommand{\twothirds}{{\ensuremath{\scriptstyle\frac{2}{3}}}}
\newcommand{\fourthirds}{{\ensuremath{\scriptstyle\frac{4}{3}}}}
\newcommand{\threehalves}{{\ensuremath{\scriptstyle\frac{3}{2}}}}

\makeatletter
\@addtoreset{equation}{section}

\makeatother

\begin{document}
\thispagestyle{empty}
\bc
\textbf{\color{blue}\large Bounds on solutions of the rotating, stratified, incompressible,\\ 
non-hydrostatic, three-dimensional Boussinesq equations}
\par\bigskip
\textbf{John D. Gibbon\footnote{j.d.gibbon@ic.ac.uk\,;~~http://www2.imperial.ac.uk/$\sim$jdg} 
and Darryl D. Holm\footnote{d.holm@ic.ac.uk\,;~~http://www2.imperial.ac.uk/$\sim$dholm}}
\par
\textbf{Department of Mathematics, Imperial College London SW7 2AZ, UK}
\ec
\par\vspace{0mm}

\begin{abstract}
We study the three-dimensional, incompressible, non-hydrostatic Boussinesq fluid equations, which are applicable to the dynamics of the 
oceans and atmosphere.  These equations describe the interplay between velocity and buoyancy in a rotating frame.  A hierarchy of dynamical variables is introduced whose members $\Omega_{m}(t)$ ($1 \leq m < \infty$) are made up from the respective sum of the $L^{2m}$-norms of vorticity and the density gradient. Each $\Omega_{m}(t)$ has a lower bound in terms of the inverse Rossby number, $Ro^{-1}$, that turns out to be crucial to the argument. For convenience, the $\Omega_{m}$ are also scaled into a new set of variables $D_{m}(t)$. By assuming the existence and uniqueness of solutions, conditional upper bounds are found on the $D_{m}(t)$ in terms of $Ro^{-1}$ and the Reynolds number $Re$. These upper bounds vary across bands in the $\{D_{1},\,D_{m}\}$ phase plane.  The boundaries of these bands depend subtly upon $Ro^{-1}$, $Re$, and the inverse Froude number $Fr^{-1}$.  For example, solutions in the lower band conditionally live in an absorbing ball in which the maximum value of $\Omega_{1}$ deviates from $Re^{3/4}$ as a function of $Ro^{-1},\,Re$ and $Fr^{-1}$. 
\end{abstract}

\vspace{2mm}
{\scriptsize\tableofcontents}

\newpage
\section{\large Introduction}\label{intro}

\subsection{\small Background}

The problem of the existence and uniqueness of solutions of the incompressible $3D$ Navier-Stokes equations remains one of the great 
open problems of modern applied mathematics \cite{Leray1934,CF1988,FMRT2001,DG1995,RRS2016,Frisch1995}. A view widely supported 
in the literature is that the addition of rotation has a regularizing effect on turbulent solutions by aligning vortices in the direction of rotation 
\cite{BFR1985,CMG1997,QCEH1997,GHK2007,SD2008}. Experiments by Taylor \cite{GIT1921}, and Staplehurst, Davidson and 
Dalziel \cite{SDD2008}, have illustrated the effect of rotation on homogeneous turbulence in tank experiments, while a summary 
of the experimental literature for rotating convection can be found in King, Stellmach and Aurnou \cite{King2012}\,: see also 
\cite{spinlab}. 

Rotation and stratification together produce additional effects beyond those of Navier-Stokes turbulence. With their inclusion, and under 
the Boussinesq approximation ($\mbox{div}\,\bu = 0$ and $\rho^{-1}\nabla p \to \rho_{0}^{-1}\nabla p$), the PDEs are called 
the incompressible, non-hydrostatic $3D$ Boussinesq equations 
\cite{Bart1995,EM1996,EM1998,BMN1996,BMNZ1997,BMN1998,BMN2000,SW2002,SL2005,Amati2005,WEHT2011,Julien2016}. 
Their fundamental role in geophysical fluid dynamics (GFD) makes the task of studying bounds on their solutions an important one. In 
this system, rotational Coriolis and buoyancy forces are available to balance the pressure forces in the horizontal and vertical directions, 
respectively. In oceanic and atmospheric dynamics, these GFD balances are effective because the acceleration is relatively small (by a factor 
of $Ro \ll 1$) compared to each of the forces in the balances\footnote{Other balances are possible in the Earth's mantle and on exoplanets\,: 
see Julien \textit{et al.} \cite{Julien2016} and references therein.}. Moreover, these balances occur at relatively large scales, much larger than 
the dissipation scale at which the nonlinearity balances the viscous force\,: see Vallis \cite{Vallis2006,Vallis2016} and Majda and Wang 
\cite{MajdaWang2006}. This is because the Reynolds number in both atmospheric and oceanic flows is $Re \sim O(10^9)$. Consequently, in 
analyses of GFD turbulence one should expect to find an interplay between the Reynolds ($Re$), Rossby ($Ro$) and Froude ($Fr$) numbers. 
The Prandtl number $Pr$ also enters our analytical considerations, but it will play a less important role than $Re$, $Ro$ and $Fr$. 

The seminal rigorous analytical results of Babin, Mahalov and Nicolaenko on periodic domains 
\cite{BMN1996,BMNZ1997,BMN1998,BMN2000}, followed by those of Chemin, Desjardins, Gallagher and Grenier \cite{Chemin2006} on 
the whole space, have demonstrated how subtle resonance effects introduced by rotation and stratification can regularize solutions of 
the $3D$ rotating, stratified, incompressible Euler and Boussinesq equations.  However, it is not the intention of this paper to re-visit 
the problem in the same manner. Instead, we assume the existence and uniqueness of solutions, and then strike out in a different direction 
to seek explicit bounds on $L^{p}$-norms of the vorticity and density gradient fields of the Boussinesq equations in terms of $Re$, $Ro^{-1}$, 
$Fr^{-1}$ and $Pr$. For the reader who does not wish to read the technical details, the paper contains a summary of results (\S\ref{summary}) 
which the reader can consult without resorting to the proofs appearing in subsequent sections and appendices. The distinctive 
dynamics investigated by Embid and Majda \cite{EM1996,EM1998}, and discussed by Wingate, Embid, Holmes-Cerfon and Taylor 
\cite{WEHT2011}, suggest that a rich variety of solutions exists. The bounds discussed here estimate how large these solutions within 
$L^{p}$-norms can become.


\subsection{\small The non-hydrostatic Boussinesq equations}

We consider the non-hydrostatic, three-dimensional Boussinesq equations\footnote{In contrast, the $3D$ Primitive equations are derived 
by adopting the hydrostatic approximation in which the vertical pressure gradient and the buoyancy force are assumed to be in balance. 
This assumption allows the third component of the material time derivative in \eqref{nse1} to be neglected. In their seminal paper, Cao and 
Titi \cite{CaoTiti2007} have shown how to prove the regularity of solutions of that system.} for a vertically stratified, incompressible flow, 
with velocity $\bu(\bx,\,t) = (u,\,v,\,w)$ ($\mbox{div}\,\bu = 0$), moving at a constant rotation about the $z$-axis  
\beq{bouss1}
\frac{D~}{Dt} &=& \partial_{t} + \bu\cdot\bnabla\,,\\
\frac{D\bu}{Dt} &+& f (\bk\times\bu)  + \rho_{0}^{-1}\rho g\bk + \rho_{0}^{-1}\nabla p = \nu \Delta\bu\label{nse1}\,,\\
\frac{D\rho}{Dt} &-& bw = \kappa\Delta\rho\label{nse2}\,.
\eeq	
The boundary conditions are taken to be periodic on a cube of volume $\mathcal{V} = [0,\,L]_{per}^{3}$. $p(\bx,\,t)$ is the pressure and 
$\rho(\bx,\,t)$ is the density fluctuation within the total density $\tilde{\rho}(\bx,\,t)$ which has been decomposed into $\tilde{\rho} = 
\rho_{0} - b z + \rho$, where $\rho_{0}$ is a constant background reference value and $b$ is the constant density gradient in the vertical 
direction.  Finally, $f$ is twice the frame rotation rate, $g$ is the acceleration due to gravity, $\nu$ is the kinematic viscosity and $\kappa$ is the 
diffusion coefficient.
\par\medskip
We non-dimensionalize equations \eqref{bouss1}--\eqref{nse2} using the following characteristic scales\,: $L$ is the length scale for 
the three spatial coordinates $\bx = (x, y, z)$, $U$ is the velocity scale and  $L/U$ is the advective time scale. Dimensionless numbers 
are the Reynolds, Rossby, Prandtl and Froude numbers defined by
\beq{nsedim1}
Re &=& \frac{LU}{\nu}\,,\qquad Ro = \frac{U}{fL}\,,\\
\label{nsedim2}Pr &=& \frac{\nu}{\kappa}\,,\qquad\quad Fr = \frac{U}{NL}\,,
\eeq
where the Brunt-V\"ais\"al\"a frequency is taken to be
\bel{Brunt}
N = (gb/\rho_{0})^{1/2}\,.
\ee 
The non-dimensionalized versions of \eqref{nse1} and \eqref{nse2} are   
\beq{nse3}
\frac{D\bu}{Dt} &+& Ro^{-1}\bk\times\bu  + Fr^{-1}\rho\,\bk + \nabla P = Re^{-1}\Delta\bu\,,\\
\frac{D\rho}{Dt} &-& Fr^{-1}w = Pr^{-1}Re^{-1}\Delta\rho\,,\label{nse4}
\eeq
where the dimensionless pressure is given by $P = p/\rho_{0}U^2$. In the standard notation, $\bom = \mbox{curl}\,\bu$ with
\bel{brotdef}
\brot = \bom + Ro^{-1}\bk\,.
\ee
Taking the curl of (\ref{nse3}) and the gradient of (\ref{nse4}) yields
\bel{nse5}
\frac{D\brot}{Dt} + Fr^{-1}\nabla\rho\times \bk = Re^{-1}\Delta\brot + \brot\cdot\nabla \bu\,,
\ee
\bel{nse6}
\frac{D\nabla\rho}{Dt} - Fr^{-1}\nabla w = Pr^{-1}Re^{-1}\Delta\nabla\rho - \nabla\bu\cdot\nabla\rho\,,
\ee
together with $\mbox{div}\,\bu = 0$. On the unit domain with periodic boundary conditions, these are 
the fundamental equations of this paper.  With the energy defined by
\bel{en1}
E(t) = \I \left(|\bu|^{2} + |\rho|^{2}\right)dV\,,
\ee
equations \eqref{nse3} and \eqref{nse4} show that 
\bel{en2}
\frac{dE}{dt} + Re^{-1}\I \left(|\bom|^{2} + Pr^{-1}|\nabla\rho|^{2}\right)dV  = 0\,.
\ee
Thus $dE/dt < 0$ and so $E(t)$ decays from its initial data $E(t) \leq E_{0}$. Thus 
\bel{Enex1}
\int_{0}^{t}\I \left(|\bom|^{2} + Pr^{-1}|\nabla\rho|^{2}\right)dV\,d\tau \leq ReE_{0}\,.
\ee
Let us now introduce a set of $L^{2m}$-norms ($1 \leq m < \infty$) of the three-dimensional vorticity field $\brot(\bx,\,t)$ 
and the density gradient $\nabla\rho$
\beq{f1a}
P_{m}(t) &=& \left(\I |\brot|^{2m}\,dV\right)^{1/2m}\,,\\
Q_{m}(t) &=& \left(\I |\nabla\rho|^{2m}\,dV\right)^{1/2m}\label{f1b}\,,
\eeq
together with their sum
\bel{Omdef}
\Omega_{m}(t) = P_{m}(t) + Q_{m}(t)\,.
\ee
In the language of GFD, $P_{m}$ estimates the shear in $L^{2m}$ and $Q_{m}$ estimates the buoyancy gradient in $L^{2m}$.
The quantity $P_{1}$ has a lower bound, namely
\beq{lb1}
P_{1}^{2}&=& \I |\brot|^{2}dV = \I \left(|\bom|^{2} + 2Ro^{-1}\bk\cdot\bom + Ro^{-2}\right)dV \geq Ro^{-2}\,.
\eeq
In equation (\ref{lb1}) the $2\bk\cdot\bom$ term drops out under integration because the volume integral of 
$\omega_{3}$ is zero on periodic boundary conditions. Using H\"older's inequality, the $\Omega_{m}$ are 
ordered such that
\beq{lb2}
Ro^{-1} &\leq& \Omega_{1} \leq \ldots \leq \Omega_{m} \ldots 
\eeq
The existence of the lower bound $Ro^{-1}$ in (\ref{lb2}) on $\Omega_{1}$, and hence on the sequence of $\Omega_{m}$, highlights 
the difference between the rotating $3D$ Boussinesq equations treated here and the non-rotating $3D$ Navier-Stokes equations, for 
which no lower bound is known to exist. This lower bound is pivotal in the proof of Lemma \ref{lem3} in \S\ref{sector2}.
Moreover, 
\bel{Enex2}
\int_{0}^{t}\Omega_{1}^{2}d\tau \leq 2\max\left(1,Pr^{-1}\right)ReE_{0} + t Ro^{-2}\,.
\ee
Clearly, a time average $\left<\cdot\right>_{T}$ is squeezed between $Ro^{-2}$ and $Ro^{-2} + 2T^{-1}\max\left(1,Pr^{-1}\right)ReE_{0}$.

\subsection{\small Summary of results}\label{summary}

For the reader who does not wish to work through the technical estimates, the main results of the paper are summarized below in 
this subsection. Proofs of these results may be found in \S\ref{techprop} and \S\ref{mainthm} and their two associated appendices.  


\subsubsection{\small Bounds on the $\Omega_{m}$}

The results of the paper involve the variables $D_{m}$ ($m=1,\,\ldots$), which section \S\ref{Omsect} shows are connected to the 
$\Omega_{m}$ by the scaling
\bel{Dmdefintro}
D_{m} = Re^{\alpha_{m}-1/2}\Omega_{m}^{\alpha_{m}}\qquad\mbox{where}\qquad \alpha_{m} = \frac{2m}{4m-3}\,.
\ee
The scaling factor $Re^{\alpha_{m}-1/2}$ in (\ref{Dmdefintro}) appears as a consequence of the factors of $Re^{-1}$ in \eqref{nse3} 
and (\ref{nse4}). We will assume that solutions exist and are unique and therefore that the $\Omega_{m}$ and $D_{m}$ are 
differentiable. Our task, based on this assumption, is to find bounds on these. 
\par\medskip
As explained in \S\ref{DmD1sect}, instead of using the set $\{D_{1}(t),\,D_{2}(t)\,, ...\,D_{m}(t)\}$, we now use the set
\bel{lamset1}
\{D_{1}(t),\,\lambda_{2}(t)\,, ...\,\lambda_{m}(t)\}\,,
\ee
which treats $D_{1}$ as the main variable and $\{\lambda_{m}(t)\}$ as a set of exponents that represent the relative 
sizes of the higher $D_{m}$ relative to $D_{1}$ through the scaling formula 
\bel{Omscal4intro}
D_{m} = C_{m}D_{1}^{A_{m,\lambda_{m}}}
\ee
where $\lambda_{m}(t)$ appears in the numerator of\,\footnote{Lemma \ref{lem3} shows how to choose the constants $C_{m}$.}
\bel{Amdefintro}
A_{m,\lambda_{m}}(t) = \frac{(m-1)\lambda_{m}(t) + 1}{4m-3}\,.
\ee
Relations among $\Omega_{m},\,\Omega_{m+1}$, $D_{m}$ and $D_{m+1}$ arise in Lemmas \ref{Dmint}, \ref{Omlem} and \ref{Dmlem} 
in \S\ref{Omsect} and the relations \eqref{Omscal4intro} and \eqref{Amdefintro} can be found in \S\ref{DmD1sect}.
\begin{figure}[htb]
\setlength{\unitlength}{5mm}
\bc
\begin{picture}(11,11)
\thicklines
\put(0,0){\vector(0,1){10}}
\put(0,0){\vector(1,0){13}}
\thicklines
\put(0,0){\line(1,1){10}}
\put(10.86,10.1){\makebox(0,0)[b]{\sf\tiny $\lambda_{m}  = 4$}}
\thinlines
\put(0,10.25){\makebox(0,0)[b]{\sf\scriptsize$D_{m}$}}
\thinlines
\put(13.5,-0.25){\makebox(0,0)[b]{\sf\scriptsize$D_{1}$}}
\thinlines
\thicklines
{\color{red}\qbezier(0,0)(6,6)(11,6.2)}
\put(12.9,6){\makebox(0,0)[b]{\sf\color{red}\tiny $\lambda_{m}  = 1+\tilde{\delta}_{3}$}}
\thicklines
{\color{red}\qbezier[25](0,0)(5.6,5.6)(10.7,5.8)}
{\color{red}\qbezier[25](0,0)(5.2,5.2)(10.5,5.4)}
\thicklines
{\color{red}\qbezier(0,0)(5.0,5.0)(10.3,5.0)}
\put(12.5,4.9){\makebox(0,0)[b]{\sf\color{red}\tiny $\lambda_{m}  = 1+\tilde{\delta}_{\infty}$}}
\put(9,8){\makebox(0,0)[b]\textbf{\sf\tiny\color{blue}Sector 2\,: Exp-decay for limited i.d.}}
\put(8.2,4.3){\makebox(0,0)[b]\textbf{\sf\tiny Sector 1\,: ball}}
\put(6,2){\makebox(0,0)[b]\textbf{\sf\tiny forbidden}}
\linethickness{.4mm}
\qbezier(0,0)(6,4)(10.2,4)
\put(11.8,3.8){\makebox(0,0)[b]{\sf\tiny $\lambda_{m}  = 1$}}
\end{picture}
\caption{\scriptsize The $\{D_{1},\,D_{m}\}$ phase plane for a chosen value of $m$ where $\lambda_{m}(t) = 1+ \delta_{m}(t)$ represents 
a phase trajectory. The multiplicity of (red) curves $\lambda_{m}  = 1 + \tildel$ represents the variation of the lower bound 
$\delta_{m}(t) \geq \tildel $ in the range $3 < m < \infty$.}\label{fig1-intro}
\ec
\end{figure}
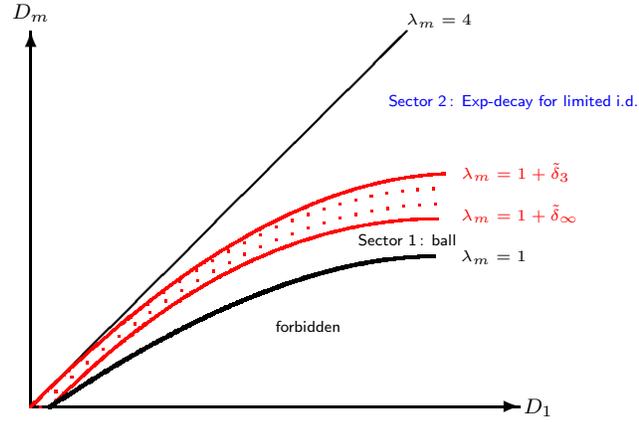
\begin{table}
\bc
\begin{tabular}{||l|l|l|c|c|l|l||}\hline
                                    & $Re$ & $Ro^{-1}$ & $Pr$ & $Fr^{-1}$ & $\tilde{\delta}_{\infty}$ & $\tilde{\delta}_{3}$ \\\hline
{\small ocean}           &  $10^{9}$       & $10^{2}$  & $10$ &  $10^{2}$  &     $6/35 = 0.171$   &  $17/35 = 0.486$ \\\hline
{\small atmosphere} &   $10^{9}$       &  $10$      &   $1$ &  $10$   &   $3/31= 0.097$  &  $13/31=0.419$\\\hline
\end{tabular}\label{table1}
\ec
\caption{\scriptsize Table of the Reynolds $Re$, inverse Rossby $Ro^{-1}$, Prandtl $Pr$ and inverse Froude $Fr^{-1}$numbers, 
together with $\tilde{\delta}_{\infty}$ and $\tilde{\delta}_{3}$ for the oceans and atmosphere of the Earth respectively.}
\end{table}
We draw a $\{D_{1},\,D_{m}\}$ phase plane (see Fig \ref{fig1-intro}) whose two main active areas are designated as sectors 1 and 2. 
Bounds in these two areas are expressed thus\,: 
\ben
\item In Fig. \ref{fig1-intro} the boundary between sectors 1 and 2 lies at $\lambda_{m} = 1 + \tildel$ where, for
\bel{RoReFr}
Ro^{-1} > 1\qquad\mbox{with}\qquad Re > 1\qquad\mbox{and}\qquad Fr^{-1} \ll Ro^{-\frac{2(m+1)}{4m+1}}Re^{\frac{3}{4m+1}}
\ee
and, in the range $3 < m < \infty$, the quantity $\tildel$ is defined by 
\beq{tildeldef}
\tildel = \frac{\ln Ro^{-1}}{\ln (ReRo^{-4/3})} + \left(\frac{3}{4m-3}\right)\frac{\ln(Re Ro^{-1})}{\ln(Re Ro^{-4/3})}\,.
\eeq
This is derived in Lemma \ref{lem3} in \S\ref{sector2}.   In \eqref{RoReFr} the range of validity on $Ro^{-1}$,  
in particular, is rather extensive. With the definitions  
\bel{del3def}
\tilde{\delta}_{\infty} = \frac{\ln Ro^{-1}}{\ln (ReRo^{-4/3})} \qquad\qquad
\tilde{\delta}_{3} = \tilde{\delta}_{\infty} + \frac{1}{3}\frac{\ln(Re Ro^{-1})}{\ln(Re Ro^{-4/3})}\,,
\ee
clearly, $\tildel$ varies between $\tilde{\delta}_{\infty}  < \tildel < \tilde{\delta}_{3}$. This range is shown 
pictorially in Fig. \ref{fig1-intro} and more specifically in Table 1. Which value of $m$ should we choose? The 
answer depends upon the value of the parameters in the range \eqref{RoReFr}, the chosen initial condition 
and, after a numerical experiment, how far the trajectory $\lambda_{m}(t)$ has travelled. The value of $m$ 
is chosen such that the region whose boundary is $\tildel$ encompasses both the initial condition and the 
path of the trajectory. In \eqref{RoReFr} if, instead, $Fr^{-1} > Ro^{-\frac{2(m+1)}{4m+1}}Re^{\frac{3}{4m+1}}$, 
then
\bel{del4def}
\tildel = \ln (Fr^{-1})/\ln(ReRo^{-4/3}) < 1\,.
\ee
In \S\ref{diffran} another range for $Fr^{-1}$ is discussed where $Ro\sim O(1)$\,: see Embid and Majda 
\cite{EM1996,EM1998}.

\item If a trajectory $\lambda_{m}(t)$ has initial conditions set in sector 1, and is assumed to remain 
in that sector, then there exists an absorbing ball for $\Omega_{1}(t)$ 
\bel{ball}
\overline{\lim}_{t\to\infty} \Omega_{1}(t)\leq \Omega_{1,rad}\,,
\ee
whose radius $\Omega_{1,rad} $ is defined by\,:
\bel{sumsec1}
\Omega_{1,rad} = c_{1,rad}Re^{\frac{3(1+ \tildel)}{4(1-\tildel)}}E_{0,Pr}^{\frac{1}{(1-\tildel)}} 
+ O\left(Ro^{-1} + Fr^{-1/2}E_{0,Pr}^{1/2} + Ro^{-1/3}E_{0,Pr}^{1/3}\right)\,,
\ee
where $E_{0,Pr} = E_{0}\max\{1,Pr\}$ and $E_{0}=E(0)$, which is derived\footnote{Strictly speaking, the radius 
of the ball is dependent upon the decaying energy $E(t)$ but here we use it's maximum.} in \S\ref{sector1}.  

\item However, if a trajectory $\lambda_{m}(t)$ has initial conditions set in sector 2 then $\Omega_{m}(t)$ 
decays exponentially provided initial data lies within the range\,:
\bel{sumsec2}
\Omega_{m}(0) \leq c_{m}Re^{\frac{3(m-1)}{4m}\left(\frac{\ln Ro^{-4/3}}{\ln(ReRo^{-4/3})}\right)} 
Ro^{-2\tildel/3\alpha_{m}}\,,
\ee
for $Ro^{-1} > 1$ and $m > 3$. This is derived in Lemma \ref{lem4} in \S\ref{sector2}. 

\item These results cannot be considered as the basis of a regularity proof for realistic values of $E_{0,Pr}$ 
and $Ro^{-1}$ because there is always the potential for a transition of a trajectory from sector 1 to sector 2, 
so we must assume that it occurs at the maximum amplitude $\Omega_{1,rad}$. This would then form the 
initial condition $\Omega_{m}(0)$ in \eqref{sumsec2} for sector 2. However, this could lie outside the range 
of $\Omega_{m}(0)$. \textbf{The only way the upper bound in \eqref{sumsec2} could reach the value of 
$\Omega_{1,rad}$ is in the limit when $Ro^{-1}$ is very large and/or $E_{0,Pr}$ is small.}
\een

\subsubsection{\small Dynamics in the ball}

The maximum amplitude of the ball for $\Omega_{1}$ is given by $\Omega_{1,rad}$, which is defined in 
\eqref{sumsec1}. If parameters are chosen such that $\tildel$, defined in \eqref{tildeldef}, is very small,  
then the departure from $Re^{3/4}$ in \eqref{sumsec1} would be minimal. However, this departure would 
increase with $\tildel$.  
The dynamics investigated by Embid and Majda \cite{EM1996,EM1998}, and discussed by Wingate \textit{et al.} 
\cite{WEHT2011}, suggest that the variety of triad solutions found in these papers would lie in this absorbing ball 
in sector 1. If $\Omega_{m}(t)$ is calculated for a given solution then the appropriate value of $\Omega_{1,rad}$ 
will be its maximum. 
\begin{figure}[htb]
\setlength{\unitlength}{5mm}
\bc
\begin{picture}(11,8)
\thicklines
{\color{red}\qbezier(0,0)(6,6)(11,6.2)}
\put(13.3,6){\makebox(0,0)[b]{\sf\color{red}\scriptsize $\lambda_{m_{0}}  = 1+\tilde{\delta}_{m_{0}}$}}
\thicklines
\put(4.5,2.4){\makebox(0,0)[b]{\sf\color{blue}\tiny i.c.}}
\put(11.7,4.7){\makebox(0,0)[b]{\sf\color{blue}\tiny Orbit~$\lambda_{m_{0}}(t)$}}
{\color{blue}\qbezier[15](4.5,3)(5,3)(10,5)}
\put(11.6,2.2){\makebox(0,0)[b]{\sf\color{red}\scriptsize $\lambda_{m_{0}}  = 1$}}
\qbezier(0,0)(4,2.5)(10.2,2.4)
\put(10.5,3.4){\makebox(0,0)[b]{\sf\color{red}\tiny sector 1~(ball)}}
\end{picture}
\caption{\scriptsize Initial conditions sector 1 (ball) in the $\{D_{1},\,D_{m}\}$ phase plane for some chosen $m=m_{0} > 3$. 
Solutions remain in the ball provided the orbit $\lambda_{m_{0}}(t)$ remains in sector 1.}\label{fig2-intro}
\ec
\end{figure}
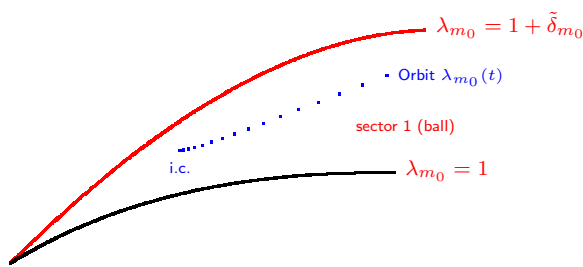
\par\medskip
Further investigation of these dynamics will require a set of numerical experiments in which, for a chosen value of $m$,  the evolution of $\Omega_{m}(t)$ is computed from an initial condition $\Omega_{m}(0)$ positioned within sector 1 (see Fig. \ref{fig2-intro}). This would involve calculating $D_{1}(0)$ and then the set $\{\lambda_{m}(0)\}$ for a chosen $m$, and then following the evolution of  $D_{1}(t)$ as well as the trajectory $\{\lambda_{m}(t)\}$, to test whether it remains in sector 1.  


\section{\large Technical properties of $\Omega_{m}$ and $D_{m}$}\label{techprop}

\subsection{\small A bounded hierarchy of time integrals of $D_{m}$}

 For the $3D$ Navier-Stokes equations, a finitely bounded hierarchy of time integrals (or averages) was proved to exist \cite{JDGCMS2011}, 
namely $\int_{0}^{t}D_{m}\,d\tau < \infty$\,: in that case the $D_{m}$ contained the vorticity alone. The proof was based on an early 
seminal paper by Foias, Guillop\'e and Temam \cite{FGT1981}. An adaptation of the same method for the Boussinesq equations, shown 
in Appendix \ref{Dmapp}, can be used to establish the same result, where $D_{m}$ is defined in \eqref{Dmdefintro}. We have 
\begin{lemma}\label{Dmint}
With $D_{m}$ defined in \eqref{Dmdefintro}, for every finite $t > 0$ and for each $1 \leq m \leq \infty$
\bel{Dmfinite}
\int_{0}^{t}D_{m}\,d\tau < \infty\,.
\ee
\end{lemma}
Note that when $m=1$, $\int_{0}^{t}\Omega_{1}^2\,d\tau < \infty$. 


\subsection{\small Differential inequalities for $\Omega_{m}$ and $D_{m}$}\label{Omsect}

The dynamical variables $\Omega_{m}$ defined in (\ref{f1a}) form the basis of the following lemma. Because the long time regularity 
of solutions is not the main issue, in the following sections we will assume that solutions exist and are unique and therefore the 
$\Omega_{m}$ are differentiable.  The technical details in the proof of this Lemma are relegated to Appendix \ref{appB} while 
Appendix \ref{appA} contains the proof of a triangular inequality involving $\Omega_{1},\,\Omega_{m}$ and $\Omega_{m+1}$. 
\begin{lemma}\label{Omlem}
For $1 < m < \infty$, the $\Omega_{m}$ obey the following differential inequality
\bel{Om3a}
\dot{\Omega}_{m} \leq \Omega_{m}\left\{- \min(Pr^{-1},1)\frac{Re^{-1}}{c_{1,m}}
\left(\frac{\Omega_{m+1}}{\Omega_{m}}\right)^{\beta_{m}} + c_{2,m}Re^{2\alpha_{m} - 1}\Omega_{m}^{2\alpha_{m}} 
+ c_{3,m}\Gamma_{m}\right\}\,.
\ee 
where
\bel{betamdef}
\alpha_{m} = \frac{2m}{4m-3}\qquad\qquad\beta_{m} = \fourthirds m(m+1)
\ee
and
\bel{Gamdef}
\Gamma_{m} = Re^{-1}+ Fr^{-1} + Ro^{-2\alpha_{m+1}} Re^{\frac{3}{4m+1}}\,.
\ee
\end{lemma}
\par\smallskip
Now define the re-scaling
\bel{Dmdef}
D_{m} = Re^{\alpha_{m}- 1/2}\Omega_{m}^{\alpha_{m}}
\ee
then (\ref{app5c}), (\ref{app6}) and (\ref{app7}) and the key formula
\bel{key}
\left(\frac{1}{\alpha_{m+1}} - \frac{1}{\alpha_{m}}\right)\beta_{m} = 2\,,
\ee
show that
\bel{Om4}
\left(\frac{\Omega_{m+1}}{\Omega_{m}}\right)^{\beta_{m}} =
Re\left(\frac{D_{m+1}}{D_{m}}\right)^{\rho_{m}} D_{m}^{2} 
\geq 2^{2m^{2}\eta_{m}}Re\left(\frac{D_{m}}{D_{1}}\right)^{\eta_{m}} D_{m}^{2}
\ee
with $\rho_{m}$ and $\eta_{m}$ given by 
\bel{rhodef}
\rho_{m} = \twothirds m(4m+1)\qquad\qquad \eta_{m} = \frac{2m}{3(m-1)}\,.
\ee
These, together with the triangular relation between $D_{1}$, $D_{m}$ and $D_{m+1}$ in (\ref{app6}),  
lead to\footnote{The factor of $2^{2m^{2}\eta_{m}}$ has been absorbed in the constant $c_{1,m}$.}
\begin{lemma}\label{Dmlem}
The $D_{m}$ defined in (\ref{Dmdef}) obey the differential inequality
\bel{Dm1}
\alpha_{m}^{-1}\dot{D}_{m} \leq D_{m}^{3}\left\{- \frac{\min(Pr^{-1},1)}{c_{1,m}}\left(\frac{D_{m}}{D_{1}}\right)^{\eta_{m}} 
+ c_{2,m}\right\} + c_{3,m}\Gamma_{m}D_{m}\,.
\ee
\end{lemma}
\textbf{Remark\,:} Apart from the value of the constants, \eqref{Dm1} is precisely the same as that for the $3D$ Navier-Stokes 
equations -- see references in \cite{JDGIMA2015}.


\subsection{The $\{D_{1},\,D_{m}\}$ relation}\label{DmD1sect}

Because the differential inequality \eqref{Dm1} in Lemma \ref{Dmlem} is identical to that derived for the $3D$ Navier-Stokes equations, we use the same method to analyze the possible dynamics. Numerical observations made in \cite{DGGKPV2013,GDGKPV2014} showed that for the $3D$ Navier-Stokes equations a scaling relation exists between $D_{m}$ and $D_{1}$ of the form\footnote{The idea was subsequently developed for $3D$ MHD in \cite{MHDsim}.} 
\bel{Omscal4}
D_{m} = C_{m}D_{1}^{A_{m,\lambda_{m}}}\,,
\ee
where $D_{m}$ contained $L^{p}$-norms of the vorticity alone. The exponents $A_{m,\lambda_{m}}$ were defined as 
\bel{Amdef}
A_{m,\lambda_{m}} = \frac{(m-1)\lambda_{m} + 1}{4m-3}\,.
\ee
In \cite{JDGIMA2015} it was shown why this is true provided $\lambda_{m} = \lambda_{m}(t)$ and $1 \leq \lambda_{m}(t) \leq 4$ 
subject to the assumption that the higher frequencies $\Omega_{m}$ for $m >1$ are explicit functions of the basic frequency 
$\Omega_{1}$ and $t$. Given the the differential inequality \eqref{Dm1} in Lemma \ref{Dmlem} is identical to that for the NSE 
(except for constants), in the light of the derivation in \cite{JDGIMA2015}, how might we justify using the same relation \eqref{Omscal4} 
for the Boussinesq equations? 
\par\smallskip
In the original variables, dimensional analysis would tell us that
\bel{k1}
\frac{\Omega_{m}}{\Omega_{1}} = \left[\kappa_{m}(t)\right]^{\frac{3(m-1)}{2m}}\,,
\ee
where $\kappa_{m}$ is a set of inverse length scales. In Appendix \ref{appA} it is shown that there exists a triangular inequality between 
$\Omega_{1}$, $\Omega_{m}$ and $\Omega_{m+1}$ such that 
\bel{k2}
\left(\frac{\Omega_{m}}{\Omega_{1}}\right)^{m^{2}} \leq 2^{m^2}\left(\frac{\Omega_{m+1}}{\Omega_{1}}\right)^{m^{2}-1}\,.
\ee
This simply translates into the ordering $\kappa_{m} \leq 2^{\eta_{m}}\kappa_{m+1}$ where $\eta_{m}$ is defined in \eqref{rhodef}.
Given the hierarchy of finite time integals in Lemma \ref{Dmint}, from \eqref{k1} we write
\bel{k4}
\int_{0}^{t}\Omega_{m}^{\alpha_{m}}\,d\tau  = \int_{0}^{t}\Omega_{1}^{\alpha_{m}}\kappa_{m}^{\frac{3(m-1)\alpha_{m}}{2m}}\,d\tau 
\leq \int_{0}^{t}\Omega_{1}^{2}\,d\tau \int_{0}^{t}\kappa_{m}\,d\tau\,.
\ee
 It can then be seen that for the right hand side of (\ref{k4}) to be finite we require
\bel{k5}
\int_{0}^{t}\kappa_{m}\,d\tau < \infty\,.
\ee
How might we choose $\kappa_{m}$ to satisfy \eqref{k5}?  The simple assumption that $\kappa_{m}$ is solely a function of $\Omega_{1}^2$ (this variable controls regularity) and $t$ allows us to choose $\kappa_{m}$ to be a \textit{concave} function of $\Omega_{1}^2$ and $t$. Jensen's inequality then ensures that \eqref{k5} is satisfied. The simplest choice is\footnote{Taking $Ro^{-1} > 1$ ensures that $\Omega_{1}^2 > 1$. The coefficient $Re^{\lambda_{m}/2}$ is freely chosen such that there is no explciti factor of $Re$ in \eqref{Omscal4}, as in the $3D$ Navier-Stokes equations.}
\bel{k6}
\kappa_{m} = c_{m}Re^{\lambda_{m}/2}\left[\Omega_{1}^{2}\right]^{\onethird(\lambda_{m}(t)-1)}\,.
\ee
Substituting \eqref{k6} into \eqref{k1} gives \eqref{Omscal4} above with the set of constants $C_{m}$  to be determined. Note that 
$\lambda_{m}(t)$ is restricted to the range 
\bel{k7}
1 \leq \lambda_{m}(t) \leq 4\,,
\ee
because of the necessity of having $\kappa_{m}$ convex in $\Omega_{1}^{2}$. 
\begin{figure}
\setlength{\unitlength}{5mm}
\bc
\begin{picture}(11,11)
\thicklines
\put(0,0){\vector(0,1){10}}
\put(0,0){\vector(1,0){13}}
\thicklines
\put(0,0){\line(1,1){10}}
\put(10.86,10.1){\makebox(0,0)[b]{\sf\scriptsize $\lambda_{m}  = 4$}}
\thinlines
\put(0,10.25){\makebox(0,0)[b]{\sf\scriptsize$D_{m}$}}
\thinlines
\put(13.5,-0.25){\makebox(0,0)[b]{\sf\scriptsize$D_{1}$}}
\thinlines
\thicklines
\put(13,6.2){\makebox(0,0)[b]{\sf\color{red}\scriptsize $\lambda_{m}  = 1+\tildel$}}
{\color{red}\qbezier(0,0)(6,6)(11,6.2)}
\put(8,7.3){\makebox(0,0)[b]\textbf{\sf\scriptsize\color{blue}Sector 2\,: Exp-decay for limited i.d.}}
\put(9.4,5){\makebox(0,0)[b]\textbf{\sf\scriptsize Sector 1\,: ball}}
\put(6,2){\makebox(0,0)[b]\textbf{\sf\scriptsize forbidden}}
\linethickness{.4mm}
\qbezier(0,0)(6,4)(11,4)
\put(13,3.8){\makebox(0,0)[b]{\sf\scriptsize $\lambda_{m}  = 1$}}
\end{picture}
\ec
\caption{\scriptsize Cartoon of the $\{D_{1},\,D_{m}\}$ phase plane ($m > 3$) parametrized by curves $\lambda_{m}(t) = {\rm const}$\,: 
no information on the direction of trajectories is known so the strategy is to estimate bounds within each sector. Sector 1 is 
bounded by the curves $\lambda_{m} =1$ and $\lambda_{m} = 1 + \tildel$. Sector 2 is bounded by $\lambda_{m} = 1 + \tildel$ 
and $\lambda_{m} = 4$. {\color{red}The restriction of $\lambda_{m}(t)$ to the range $1 \leq \lambda_{m}(t) \leq 4$ occurs 
because of the necessity of having $\kappa_{m}$ convex in \eqref{k6}. At $\lambda_{m}=4$ $A_{m,4}=1$ so $D_{m} = 
C_{m}D_{1}$, which is the straightline.} Sector 3 lying above $\lambda_{m} = 4$, is a highly pathological region.}\label{cart1}
\end{figure}
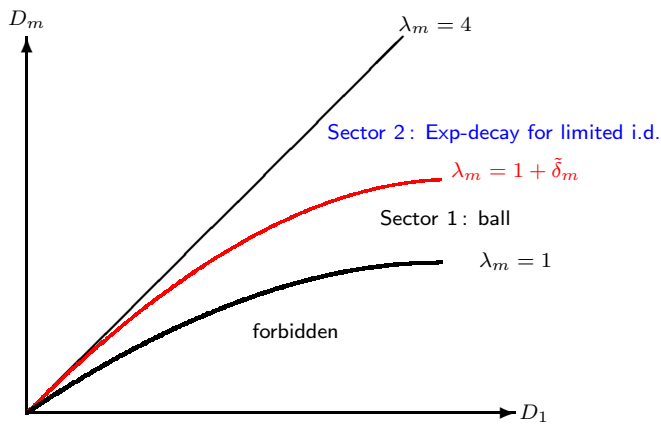


\section{\large Proof of the main results}\label{mainthm}

In the following subsections the constants $\tilde{c}_{i,m}$ depend on the $c_{i,m}$  $(i=1,2,3)$ and are defined by\,:
\bel{consts1}
\tilde{c}_{1,m} = 2\alpha_{m}\min(Pr^{-1},1)c_{1,m}^{-1}\,,\qquad\qquad\tilde{c}_{2,m} = 2\alpha_{m}c_{2,m}\qquad
\qquad\tilde{c}_{3,m} = 2\alpha_{m}c_{3,m}\,.
\ee


\subsection{\small Proof for Sector 2}\label{sector2} 

\begin{lemma}\label{lem3} (sector 2) For $m > 3$ with $C_{m}$ chosen such that $\tilde{c}_{1,m}\tilde{C}_{m} > 2\tilde{c}_{2,m}$, 
let $\tilde{\delta}_{m}$ be defined by
\bel{deltatildedef}
\tildel = \frac{\ln Ro^{-1}}{\ln (ReRo^{-4/3})} + \left(\frac{3}{4m-3}\right)\frac{\ln(Re Ro^{-1})}{\ln(Re Ro^{-4/3})}\,.
\ee
If $Ro$, $Re$ and $Fr$ satisfy
\bel{Robd1}
Ro^{-1} > 1\,,\quad Re > 1\,,\qquad\mbox{and}\qquad Fr^{-1} \ll Ro^{-\frac{2(m+1)}{4m+1}}Re^{\frac{3}{4m+1}}\,,
\ee
and initial data $D_{m}(0)$ satisfies 
\bel{DmidB}
D_{m}(0) \leq \left[\shalf\tilde{c}_{2,m}^{-1}\tilde{c}_{1,m}\tilde{C}_{m}\right]^{1/2}
Re^{\shalf\tildel} Ro^{-\twothirds\tildel}\,,
\ee
which, in terms of $\Omega_{m}$, is
\bel{trans5}
\Omega_{m}(0) \leq \left[\shalf\tilde{c}_{2,m}^{-1}\tilde{c}_{1,m}\tilde{C}_{m}\right]^{1/2\alpha_{m}}
Re^{\frac{2-\alpha_{m}}{4\alpha_{m}}\left(\frac{\ln Ro^{-4/3}}{\ln(ReRo^{-4/3})}\right)} 
Ro^{-2\tildel/3\alpha_{m}}\,,
\ee
then when $\lambda_{m}(t)$ lies in the range
\bel{lem3A}
1+ \tilde{\delta}_{m}\leq \lambda_{m}(t) \leq 4
\ee
then $D_{m}(t)$ decays exponentially in time.  
\end{lemma}
\textbf{Note\,:} The exponent of $Re$ in the estimate in \eqref{trans5} for $\Omega_{m}(0)$ is small unless $Ro^{-1}$ is large.
\par\medskip\noindent
\textbf{Proof\,:} Lemma \ref{Dmlem} shows that the $D_{m}$ satisfy the differential inequality (\ref{Dm1}) when 
$1 < m < \infty$, where $\eta_{m}$ is defined in \eqref{rhodef}, and the constants satisfy $c_{1,m}^{-1} < c_{2,m}$ with $\Gamma_{m}$ 
is defined as in \eqref{Gamdef}. Dividing \eqref{Dm1} by $D_{m}^{3}$ gives 
\bel{Dmdi2}
\frac{d~}{dt} D_{m}^{-2} \geq X_{m}(t)D_{m}^{-2} - \tilde{c}_{2,m}\,.
\ee
$X_{m}(t)$ is defined as 
\bel{Xmdef}
X_{m} = \tilde{c}_{1,m}D_{m}^{2}\left(\frac{D_{m}}{D_{1}}\right)^{\eta_{m}} - \tilde{c}_{3,m}\Gamma_{m}\,.
\ee
A positive lower bound is required on the time integral of $X_{m}(\tau)$ to show that $D_{m}^{-2}(t)$ 
never passes through zero for some range of initial conditions. To achieve this we introduce the relation 
between $D_{m}$ and $D_{1}$ in (\ref{Omscal4}) and then use the lower bound on $D_{1}$ in (\ref{lb2}) 
and (\ref{Dmdef}). The result turns out to be
\bel{lbD1}
D_{1}(t) \geq Re^{3/2}Ro^{-2}\,.
\ee
Noting that 
\bel{etarel}
\eta_{m} + 2 = 2(4m-3)/3(m-1) = 2\eta_{m}\alpha_{m}^{-1}
\ee
 it is found that ($\tilde{C}_{m} = C_{m}^{2+\eta_{m}}$)
\bel{Xmint1}
X_{m}(t) =  \tilde{c}_{1,m}\tilde{C}_{m} D_{1}^{2(\lambda_{m} -1)/3} - \tilde{c}_{3,m}\Gamma_{m}\,.
\ee
Using (\ref{lbD1}) with a designated lower bound on $\lambda_{m}(t)$ as $\lambda_{m} \geq 1+\tildel$, the lower bound 
on $X_{m}$ is $\mathcal{R}_{m,\tildel}$, which is defined as
\bel{Rmdef}
\mathcal{R}_{m,\tildel} =  \tilde{c}_{1,m}\tilde{C}_{m}Re^{\tildel}Ro^{-4\tildel/3} -  \tilde{c}_{3,m}\Gamma_{m}\,, 
\ee
so that
\bel{lbXm}
X_{m}(t) \geq \mathcal{R}_{m,\tildel}\,.
\ee
To be sure that $\mathcal{R}_{m,\tildel} > 0$, we need 
\bel{Rm1}
\tilde{c}_{1,m}\tilde{C}_{m}Re^{\tildel}Ro^{-4\tildel/3} > 
\tilde{c}_{3,m}\left(Re^{-1}+ Fr^{-1} + Ro^{-2\alpha_{m+1}} Re^{\frac{3}{4m+1}}\right)\,.
\ee
For $Re \gg 1$, the first term on the right $Re^{-1}$ may be regarded as negligible, and the $Ro^{-2\alpha_{m+1}}$-term 
dominates over $Fr^{-1}$, which, according to the stated regime, we neglect. We write (\ref{Rm1}) as
\bel{Rm2}
\tilde{c}_{1,m}\tilde{C}_{m}(Re Ro^{-4/3})^{\left(\tildel - \threehalves\alpha_{m+1}\right)} > 
\tilde{c}_{3,m}Re^{\frac{3}{4m+1} - \threehalves\alpha_{m+1}} 
\ee
where $\frac{3}{4m+1} - \threehalves\alpha_{m+1} = -\frac{3m}{4m+1} $. To satisfy (\ref{Rm2}) we make our first choice of 
$\tildel$ as
\bel{Rm3}
\twothirds\tildel^{(1)} = \alpha_{m+1} - \frac{2m}{4m+1}\left(\frac{\ln Re}{\ln(ReRo^{-4/3})}\right)\,,
\ee
and choose $\tilde{C}_{m}$ such that $\tilde{C}_{m} > \tilde{c}_{3,m}\tilde{c}_{1,m}^{-1}$. After a little algebra, \eqref{Rm3} 
becomes
\bel{Rm3A}
\tildel^{(1)} = \frac{\ln Ro^{-1}}{\ln (ReRo^{-4/3})} + \left(\frac{3}{4m+1}\right)\frac{\ln(Re Ro^{-1})}{\ln(Re Ro^{-4/3})}\,,
\ee
where, of course, $Ro^{-1} > 1$ and $Re > 1$.
\par\vspace{4mm}
Now we turn to the differential inequality (\ref{Dmdi2}) and use the fact that 
\beq{Xmint2}
\int_{0}^{t}X_{m}(t')\,dt' &\geq&  t\,\mathcal{R}_{m,\tildel}\,.
\eeq
This allows us to integrate (\ref{Dmdi2}) to obtain
\beq{Xmint3}
\shalf [D_{m}(t)]^{2} &\leq& \frac{\exp\{-\int_{0}^{t}X_{m}(t')\,dt'\}}{\frac{1}{2}[D_{m}(0)]^{-2} 
- \tilde{c}_{2,m}\int_{0}^{t}\exp\{- \int_{0}^{t'}X_{m}(t'')\,dt''\}dt'}\\
&=& \frac{\exp\{-c_{m}\tilde{c}_{1,m}t\,\mathcal{R}_{m,\tildel}\}}
{\shalf [D_{m}(0)]^{-2} - \tilde{c}_{2,m}\left[\mathcal{R}_{m,\tildel}\right]^{-1}
\left(1 - \exp\{-t\,\mathcal{R}_{m,\tildel}\}\right)}\,.\nonumber
\eeq
The denominator of inequality (\ref{Xmint3}) cannot develop a zero if initially
\beq{id1}
D_{m}(0) &\leq& \left[\shalf\tilde{c}_{2,m}^{-1}\mathcal{R}_{m,\tildel}\right]^{1/2}\,,
\eeq
in which case this part of the solution decays exponentially.  Given that $Ro^{-1}$ acts as a lower bound on the 
sequence of $\Omega_{m}$, (\ref{Dmdef}) shows that $D_{m}(0)$ must sit in the annular region
\bel{ann1}
Re^{\alpha_{m}-1/2}Ro^{-\alpha_{m}} \leq D_{m}(0) \leq \left[\shalf\tilde{c}_{2,m}^{-1}\mathcal{R}_{m,\tildel}\right]^{1/2}\,.
\ee
From the definition of $\mathcal{R}_{m,\tildel}$ in (\ref{Rmdef}) we need to satisfy
\bel{Rdel1}
Re^{\alpha_{m}-1/2}Ro^{-\alpha_{m}} <  \left[\shalf\tilde{c}_{2,m}^{-1}\tilde{c}_{1,m}\tilde{C}_{m}\right]^{1/2}
Re^{\tildel/2}Ro^{-2\tildel/3}\,,
\ee
where we make our second choice of $\tildel$ as
\bel{delta3}
\twothirds\tildel^{(2)}  = \alpha_{m} - \twothirds(1-\shalf\alpha_{m})\frac{\ln Re}{\ln (ReRo^{-4/3})}\,,
\ee
with $\tilde{c}_{1,m}\tilde{C}_{m} > 2\tilde{c}_{2,m}$.  \eqref{delta3} can be rewritten as
\bel{delta3A}
\tildel^{(2)} = \frac{\ln Ro^{-1}}{\ln (ReRo^{-4/3})} + \left(\frac{3}{4m-3}\right)\frac{\ln(Re Ro^{-1})}{\ln(Re Ro^{-4/3})}\,.
\ee
Given that  $\tildel^{(1)}$ and $\tildel^{(2)}$ in (\ref{Rm3A}) and (\ref{delta3A}) are not the same, we need to choose 
the larger of the two. Clearly $\tildel^{(2)}$ is the larger of the two. We also require $\tildel^{(1)} > 0$ and 
$\tildel^{(2)} < 1$. The first is true  provided $Re > 1$ and $Ro^{-1} > 1$. The second requires $m > 3$. 
\par\medskip
Finally we address the case where $Fr^{-1} > Ro^{-2\alpha_{m+1}} Re^{\frac{3}{4m+1}}$ and is thus the 
dominant term in \eqref{Rm1}. $\tildel$ comes out to be 
\bel{Frdelta1}\tildel = \frac{\ln Fr^{-1}}{\ln(ReRo^{-4/3})}
\ee
where the only restriction to be imposed on $Fr^{-1}$ is that $\tildel < 1$. \hfil $\blacksquare$


\subsection{\small Proof for Sector 1}\label{sector1}

\begin{lemma}\label{lem4} (sector 1) When $\lambda_{m}(t)$ lies in the range $ 1 \leq \lambda_{m}(t) \leq 1 + \tildel$ there 
exists an absorbing ball for $D_{1}(t)$ of radius $D_{1} \leq D_{1,rad}^{(1)}$ which is defined by
\beq{D1raddef}
D^{(1)}_{1,rad} &=& c\,\left(Re^{3/2}E_{0,Pr}\right)^{\frac{2}{1-\tildel}} + Ro^{-2}Re^{3/2} + Re^{3/2}O\left(Fr^{-1}E_{0,Pr} 
+ Ro^{-2/3}E_{0,Pr}^{2/3}\right)
\eeq
or, expressed in terms of terms of $\Omega_{1,rad}$,
\bel{trans2}
\Omega_{1,rad} = c\,Re^{\frac{3(1+ \tildel)}{4(1-\tildel)}}E_{0,Pr}^{\frac{1}{(1-\tildel)}} + O\left(Ro^{-2} + Fr^{-1}E_{0,Pr} 
+ Ro^{-2/3}E_{0,Pr}^{2/3}\right)^{1/2}\,.
\ee
\end{lemma}
\par\vspace{3mm}\noindent
\textbf{Proof\,:} From the definition $D_{m} = Re^{\alpha_{m} - 1/2}\Omega_{m}^{\alpha_{m}}$, a formal differential inequality for 
\bel{D1ab}
D_{1} = Re^{3/2}\Omega_{1}^{2} = Re^{3/2}(P_{1}+Q_{1})^{2}
\ee
is
\beq{D1a}
\shalf\dot{D}_{1} = Re^{3/2}\Omega_{1}\dot{\Omega}_{1} &=& Re^{3/2} (P_{1}+Q_{1})(\dot{P}_{1}+\dot{Q}_{1})\non\\
&=& \shalf Re^{3/2}\partial_{t}(P_{1}^{2} + Q_{1}^{2}) + Re^{3/2}(Q_{1}\dot{P}_{1}+P_{1}\dot{Q}_{1}) 
\eeq
We write 
\bel{D1b}
\shalf \partial_{t}(P_{1}^{2}) \leq - Re^{-1}\I |\nabla\bom|^{2}dV + N_{P}\qquad 
\qquad \shalf \partial_{t}(Q_{1}^{2}) \leq - Re^{-1}Pr^{-1}\I |\Delta\rho|^{2}dV + N_{Q}
\ee
where
\beq{NpNqdef}
N_{P} &=&  \I |\nabla\bu||\brot|^{2}dV + Fr^{-1}\I |\brot||\nabla\rho|\,dV\non\\
N_{Q} &=& \I |\nabla\bu||\nabla\rho|^{2}dV + Fr^{-1}\I |\nabla\rho||\nabla w|\,dV\,.
\eeq
A formal differential inequality for $D_{1}$ thus becomes
\beq{D1c}
\shalf\dot{D}_{1} &\leq& - Re^{1/2}\I \left(|\nabla\bom|^{2} + Pr^{-1}|\Delta\rho|^{2}\right)\,dV + 
+ Re^{3/2}(Q_{1}\dot{P}_{1}+P_{1}\dot{Q}_{1}) + Re^{3/2}(N_{P}+N_{Q}) \non\\
&=& - Re^{1/2}(1 + 2Q_{1}/P_{1})\I |\nabla\bom|^{2}dV - Re^{1/2}Pr^{-1}(1 + 2P_{1}/Q_{1})\I |\Delta\rho|^{2}\,dV\non\\
& + & 2Re^{3/2}\left(Q_{1} N_{P}/P_{1} + P_{1} N_{Q}/Q_{1}\right)\,.
\eeq
We also have
\beq{D1d}
N_{P} &=&\I |\brot|^{\frac{2m-3}{m-1}}|\brot|^{\frac{1}{m-1}}|\nabla\bu|dV + 
Fr^{-1} \I |\brot||\nabla\rho|\,dV \non\\
&\leq& P_{1}^{\frac{2m-3}{m-1}}P_{m}^{\frac{1}{m-1}}\|\nabla\bu\|_{2m} + Fr^{-1}P_{1}Q_{1} \,.
\eeq
Likewise
\beq{D1e}
N_{Q} &=&\I |\nabla\rho|^{\frac{2m-3}{m-1}}|\nabla\rho|^{\frac{1}{m-1}}|\nabla\bu|dV
+ Fr^{-1}\I |\nabla\rho||\nabla w|\,dV\non\\
&\leq& Q_{1}^{\frac{2m-3}{m-1}}Q_{m}^{\frac{1}{m-1}}\|\nabla\bu\|_{2m} + Fr^{-1}\left(P_{1} + Ro^{-1}\right)Q_{1} \,.
\eeq
where account has been taken of the difference between $\bom$ and $\brot$. 
Therefore, for $m \geq 2$, the last term in (\ref{D1c}) is estimated as 
\beq{D1f}
Q_{1} N_{P}/P_{1} + P_{1} N_{Q}/Q_{1} \leq 2 \Omega_{1}^{\frac{2m-3}{m-1}}\Omega_{m}^{\frac{1}{m-1}}\|\nabla\bu\|_{2m}
+ Fr^{-1}\left(\Omega_{1}^{2} + Ro^{-1}\Omega_{1}\right)\,. 
\eeq
In the last line we can use $\|\nabla\bu\|_{2m} \leq c_{m} \|\bom\|_{2m}$, for $1 < m < \infty$, but this estimate is 
in terms of $\bom$ and not $\brot$ so it is necessary to make an adjustment by using $\bom = \brot - Ro^{-1}$, 
\beq{D1g}
Q_{1} N_{P}/P_{1} + P_{1} N_{Q}/Q_{1} &\leq& c_{m}\left[Re^{-3/2}D_{1}^{\frac{2m-3}{2(m-1)}}D_{m}^{\frac{4m-3}{2(m-1)}} + 
Ro^{-1}Re^{-\frac{3(2m-1)}{4m}}D_{1}^{\frac{2m-3}{2(m-1)}}D_{m}^{\frac{4m-3}{2m(m-1)}}\right]\non\\
&+& Fr^{-1}\left(Re^{-3/2}D_{1} + Re^{-3/4}Ro^{-1}D_{1}^{1/2}\right)\,.
\eeq
Inserting the relation between $D_{m}$ and $D_{1}$ from (\ref{Omscal4})
\bel{DmD1a}
D_{m} = C_{m}D_{1}^{A_{m,\lambda_{m}}}\qquad\qquad 
A_{m,\lambda_{m}} = \frac{(m-1)\lambda_{m} + 1}{4m-3}
\ee
transforms (\ref{D1g}) into 
\beq{D1h}
Q_{1} N_{P}/P_{1} + P_{1} N_{Q}/Q_{1}  &\leq& c_{2,m}\left\{Re^{-3/2}D_{1}^{\xi_{m,\lambda_{m}}}
+ Ro^{-1}Re^{-\frac{3(2m-1)}{4m}} D_{1}^{\chi_{m,\lambda_{m}}}\right\}\non\\
&+& Fr^{-1}\left(Re^{-3/2}D_{1} + Re^{-3/4}Ro^{-1}D_{1}^{1/2}\right)\,,
\eeq
where $\xi_{m,\lambda_{m}}$  is defined as 
\beq{D1i}
\xi_{m,\lambda_{m}}(\tau) &=& \frac{A_{m,\lambda_{m}}(4m-3) +2m - 3}{2(m-1)}\non\\
&=& 1 + \shalf \lambda_{m} = (3+ \delta_{m})/2
\eeq
and $\chi_{m,\lambda_{m}}$ as
\bel{chidef}
\chi_{m,\lambda_{m}} = 1 + \frac{\lambda_{m}-1}{2m} = 1+ \frac{\delta_{m}}{2m}\,.
\ee
The negative terms in (\ref{D1c}) are dealt with using the fact that $(1+Q_{1}/P_{1}) \geq 1$ and $(1+P_{1}/Q_{1}) \geq 1$ 
thus giving
\beq{D1j}
\shalf\dot{D}_{1} &\leq& - Re^{1/2}\I \left(|\nabla\bom|^{2} + Pr^{-1}|\Delta\rho|^{2}\right)\,dV + 
c_{2,m}\left\{Re^{-3/2}D_{1}^{\xi_{m,\lambda_{m}}} 
+ Ro^{-1}Re^{-\frac{3(2m-1)}{4m}}D_{1}^{\chi_{m,\lambda_{m}}}\right\}\non\\
&+& Fr^{-1}\left(Re^{-3/2}D_{1} + Re^{-3/4}Ro^{-1}D_{1}^{1/2}\right)\,.
\eeq
Dealing with the negative terms in (\ref{D1j}) first, an integration by parts gives
\beq{D1k}
\I |\brot|^{2}\,dV &=& \I |\bom|^{2}\,dV + Ro^{-2}\non\\
&\leq& \left(\I|\nabla\bom|^{2}dV\right)^{1/2}\left(\I|\bu|^{2}dV\right)^{1/2} + Ro^{-2}\,,
\eeq
and similarly for $\I |\nabla\rho|^{2}\,dV$ without the lower bound. The energy in dimensionless form 
$E = \I\left(|\bu|^{2} + |\rho|^{2}\right)\,dV$, is always bounded such that $E \leq E_{0}$. Because 
\bel{D1kA}
\I |\brot|^{2}\,dV  - Ro^{-2} > 0
\ee
we can write an inequality with a positive left hand side
\bel{D1kB}
\I (|\brot|^{2} + |\nabla\rho|^{2})\,dV - Ro^{-2} \leq \sqrt{2}
E_{0,Pr}^{1/2}\left[\I\left(|\nabla\bom|^{2} + Pr^{-1}|\Delta\rho|^{2}\right)dV\right]^{1/2} 
\ee
where $E_{0,Pr} = E_{0}\max\{1,\,Pr\}$.  On squaring and losing and $Ro^{-4}$-term, becomes 
\bel{D1kC}
\left(\I (|\brot|^{2} + |\nabla\rho|^{2})\,dV \right)^{2} - 2 Ro^{-2}\left( \I (|\brot|^{2} + |\nabla\rho|^{2})\,dV \right) \leq 
2 E_{0,Pr}\I\left(|\nabla\bom|^{2} + Pr^{-1}|\Delta\rho|^{2}\right)dV
\ee
or
\bel{D1l}
\I\left(|\nabla\bom|^{2} + |\Delta\rho|^{2}\right)dV 
\geq \shalf E_{0,Pr}^{-1}Re^{-3}D_{1}^{2} - Ro^{-2}Re^{-3/2}D_{1}E_{0,Pr}^{-1}\,.
\ee
Finally, for $m \geq 3$, with $0 < \delta_{m}(t) < \tilde{\delta}_{m}$, we have 
\beq{D1m}
\shalf\dot{D}_{1} &\leq& - \shalf D_{1}^{2}/(Re^{3}E_{0})  + c_{2,m}\left\{Re^{-3/2}D_{1}^{(3 + \tilde{\delta}_{m})/2} 
+ Ro^{-1}Re^{-\frac{3(2m-1)}{4m}}D_{1}^{1 + \frac{\tilde{\delta}_{m}}{2m}}\right\}\non\\ 
&+&  Fr^{-1}\left(Re^{-3/2}D_{1} + Re^{-3/4}Ro^{-1}D_{1}^{1/2}\right) + \shalf Ro^{-2}Re^{-3/2}D_{1}E_{0}^{-1}\,.
\eeq
Therefore $D_{1}$ is controlled by a ball of radius $D^{(1)}_{m,rad}$ given by
\bel{D1raddefA}
D^{(1)}_{1,rad} = c\,\left(Re^{3/2}E_{0,Pr}\right)^{\frac{2}{1-\tildel}} + Ro^{-2}Re^{3/2} + O\left(Fr^{-1}Re^{3/2}E_{0,Pr} 
+ Ro^{-2/3}Re^{3/2}E_{0,Pr}^{2/3}\right)\,.
\ee
which is also given in (\ref{D1raddef}). This ends the proof. \hfil $\blacksquare$


\subsection{\small A different range for $Fr^{-1}$}\label{diffran}

There is another limit to one made in \S\ref{sector2}, where $Fr^{-1}$ has been neglected. The definition of $\mathcal{R}_{m,\tildel}$, 
given in (\ref{Rmdef}), and repeated here
\bel{RmdefA}
\mathcal{R}_{m,\tildel} =  \tilde{c}_{1,m}\tilde{C}_{m} Re^{\tildel}Ro^{-4\delta_{m}/3} -  \tilde{c}_{3,m}
\left(Re^{-1}+ Fr^{-1} + Ro^{-2\alpha_{m+1}} Re^{\frac{3}{4m+1}}\right)\,,
\ee
is the key to the issue. We must choose $\tilde{C}_{m}$ and $\tildel$ such that $\mathcal{R}_{m,\tildel} > 0$ but now in the regime
\bel{2nd case}
Ro\sim O(1)\qquad\mbox{with}\qquad Re^{\frac{3}{4m+1}} < Fr^{-1} < Re^{\tildel}\,,
\ee
(see Embid and Majda \cite{EM1996,EM1998}). We find $\tilde{\delta}_{m}$ subject to the constraint of (\ref{Rdel1}). In fact the choices 
of the constants are identical to the previous case but where we must now choose $\tildel$ as 
\bel{2nddeltatilde}
\tildel  = 2\alpha_{m}-1 = \frac{3}{4m-3}\,,
\ee
subject to $0 < \tildel < 1$. This means that $m > \threehalves$. Thus we have\,: 
\begin{lemma}\label{rotNSthm2}
Let $m > \threehalves$. Moreover, let $Ro \sim O(1)$ and let $Fr^{-1}$ lie in the range
\bel{Robd2}
Re^{\frac{3}{4m+1}} \ll Fr^{-1} \ll Re^{\frac{3}{4m-3}}\,.
\ee
In addition, let the set of constants $C_{m}$ in (\ref{Omscal4}) must be chosen as in the previous lemmas. 
\par\medskip\noindent
For a trajectory $\lambda_{m}(t) \geq 1$ moving through the $D_{1}-D_{m}$ plane\,:
\ben\itemsep -1mm
\item \textbf{Sector 1\,:} When $\lambda_{m}(t)$ lies in the range $1 \leq \lambda_{m}(t) \leq 1+ \tilde{\delta}_{m}$ 
then there exists an absorbing ball for $D_{1}(t)$ of radius $D^{(1)}_{1,rad}$\,; 

\item \textbf{Sector 2\,:} When $\lambda_{m}(t)$ lies in the range $1+ \tilde{\delta}_{m} \leq \lambda_{m}(t) \leq 4$ and initial data 
$D_{m}(0)$ satisfies
\bel{DmidA}
D_{m}(0) \leq \left[\shalf\tilde{c}_{2,m}^{-1}\tilde{c}_{1,m}\tilde{C}_{m}\right]^{1/2} Re^{\shalf\tildel}\,,
\ee
then $D_{m}(t)$ decays exponentially in time.  
\een
\end{lemma} 
\par\medskip\noindent
\textbf{Acknowledgements\,:} We wish to thank Professor Beth Wingate of the University of Exeter for a helpful and 
constructive reading of the manuscript. We are grateful for expressions of interest in this work shown by our friends 
at the EPSRC Centre for Doctoral Training program in Mathematics of Planet Earth, run jointly by Imperial College London 
and the University of Reading. Work by DDH is partially supported by the ERC Advanced Grant 26732 FCCA and the 
EPSRC Standard Grant EP/N023781/1.


\appendix

\section{\large The proof of Lemma \ref{Dmint}}\label{Dmapp}

Let us consider equations \eqref{nse3} and \eqref{nse4} with $Re$, $Ro$ and $Pr$ set to unity for ease of calculation. Let
\bel{HKdef}
I_{n} = \I \left(|\nabla^{n}\bu|^{2} + |\nabla^{n}\rho|^{2}\right)\,dV\,.
\ee
The essential core of these results were first worked out in \cite{FGT1981} but here we only sketch the proof following the methods 
shown in \cite{DG1995}. By differentiating both equations \eqref{nse3} and \eqref{nse4} $n$ times, using Leibnitz' theorem and 
the Gagliardo-Nirenberg inequalities, it is easy to show that 
\bel{Ina}
\shalf \dot{I}_{n} \leq -\quart I_{n+1} + c_{n,1}\left(\|\bu\|_{\infty}^{2} + \|\rho\|_{\infty}^{2} \right) I_{n} +  c_{n,2}I_{n}\,,
\ee
where $c_{n,1}$ and $ c_{n,2}$ are generic $n$-dependent constants. Now we use a version of Agmon's inequality for $n \geq 2$
\bel{In2}
\|\bu\|_{\infty}\leq c_{n}\|\nabla^{n}\bu\|_{2}^{\frac{1}{2(n-1)}} \|\nabla\bu\|_{2}^{\frac{2n-3}{2(n-1)}} \,,
\ee
and likewise for $\|\rho\|_{\infty}$, to obtain
\bel{In3}
\shalf \dot{I}_{n} \leq -\quart I_{n+1} + c_{n,1} I_{n}^{\frac{2n-1}{2(n-1)}}I_{1}^{\frac{2n-3}{2(n-1)}} + c_{n,2}I_{n}\,.
\ee
Dividing by $I_{n}^{\frac{2n}{2n-1}}$ and integrating with respect to time, we find
\bel{In4}
\quart \int_{0}^{t}\frac{I_{n+1}}{I_{n}^{\frac{2n}{2n-1}}} \,d\tau 
\leq c_{n,1} \left(\int_{0}^{t}\tilde{\kappa}_{n,1}\,d\tau \right)^{\frac{1}{2n-1}}
\left(\int_{0}^{t}I_{1}\,d\tau \right)^{\frac{2(n-1)}{2n-1}} + c_{n,2}I_{n}^{\frac{-1}{2n-1}}\,,
\ee
where $\tilde{\kappa}_{n,1}$ is defined as
\bel{In5}
\tilde{\kappa}_{n,1} = \left(\frac{I_{n}}{I_{1}}\right)^{\frac{1}{2(n-1)}}\,.
\ee
Thus we have
\beq{In6}
\int_{0}^{t}\tilde{\kappa}_{n+1,1} \,d\tau &=& \int_{0}^{t}\left(\frac{I_{n+1}}{I_{n}^{\frac{2n}{2n-1}}}\right)^{1/2n}
\tilde{\kappa}_{n,1}^{\frac{2(n-1)}{2n-1}}I_{1}^{\frac{1}{2n(2n-1)}}\,d\tau\nonumber\\
&\leq& \left(\int_{0}^{t}\frac{I_{n+1}}{I_{n}^{\frac{2n}{2n-1}}}d\tau\right)^{1/2n}
\left(\int_{0}^{t}\tilde{\kappa}_{n,1}\,d\tau\right)^{\frac{2(n-1)}{2n-1}}
\left(\int_{0}^{t}I_{1}\,d\tau\right)^{\frac{1}{2n(2n-1)}}\nonumber\\
&\leq& \left(\int_{0}^{t}\tilde{\kappa}_{n,1}\,d\tau\right)^{\frac{4n^{2}-4n+1}{2n(2n-1)}}
\left(\int_{0}^{t}I_{1}\,d\tau\right)^{\frac{1}{2n}}\nonumber\\
&\leq& \frac{4n^{2}-4n+1}{2n(2n-1)}\left(\int_{0}^{t}\tilde{\kappa}_{n,1}\,d\tau\right) + \frac{1}{2n}\left(\int_{0}^{t}I_{1}\,d\tau\right)
\eeq
Thus we have a recursion relation for $n \geq 2$. To  begin the sequence we need to estimate $\int_{0}^{t}\tilde{\kappa}_{2,1}\,d\tau$. 
Using Agmon's inquality for $n=2$ and then a H\"older inequality, we can write
\bel{In7}
\shalf \dot{I}_{1} \leq - \quart I_{2} + c\,I_{1}^{3}\,.
\ee
Then
\bel{In8}
\int_{0}^{t} \frac{I_{2}}{I_{1}^2}\,d\tau \leq c\,\int_{0}^{t}I_{1}\,d\tau + O\left(I_{1}^{-1}\right)
\ee
in which case
\beq{In9}
\int_{0}^{t}\tilde{\kappa}_{2,1}\,d\tau = \int_{0}^{t} \left(\frac{I_{2}}{I_{1}}\right)^{1/2} d\tau
= \int_{0}^{t} \left(\frac{I_{2}}{I_{1}^2}\right)^{1/2} I_{1}^{1/2}\,d\tau
\leq c\,\int_{0}^{t}I_{1}\,d\tau\,.
\eeq
Thus, starting at $n=2$ the right hand side of \eqref{In6} is dependent upon $\int_{0}^{t}I_{1}\,d\tau$ which, from \eqref{en2}, we know to be 
bounded for every finite $t>0$. In consequence, we have
\bel{ln10}
\int_{0}^{t}\tilde{\kappa}_{n,1}\,d\tau < \infty\,,\qquad\qquad 
\int_{0}^{t}\|\bu\|_{\infty}\,d\tau < \infty\,,\qquad\qquad 
\int_{0}^{t} I_{n}^{\frac{1}{2n-1}}\,d\tau < \infty\,.
\ee
The last two of these three were proved by Foias, Guillop\'e and Temam \cite{FGT1981} for the $3D$ Navier-Stokes equations. 
\par\smallskip
The final step uses the Gagliado-Nirenberg inequality
\bel{ln11}
\|\nabla\bu\|_{2} \leq c_{n,m} \|\nabla^{n}\bu\|^{a}_{2}\|\nabla\bu\|_{2}^{1-a}\,,\qquad\qquad a= \frac{3(m-1)}{2m(n-1)}\,,
\ee
to write 
\bel{ln12}
\Omega_{m} \leq c_{n,m} I_{n}^{a/2}I_{1}^{(1-a)/2} 
\ee
so, raising $\Omega_{m}$ to the power $\alpha$, which is to be determined, we write
\bel{In13}
\int_{0}^{t} \Omega_{m}^{\alpha}\,d\tau \leq c_{n,m} \int_{0}^{t} I_{n}^{\alpha a/2}I_{1}^{\alpha(1-a)/2}\,d\tau\,.
\ee
We now exploit the third result in (\ref{ln10}) to write this as 
\beq{In14}
\int_{0}^{t} \Omega_{m}^{\alpha}\,d\tau &\leq& c_{n,m} \int_{0}^{t} \left[I_{n}^{\frac{1}{2n-1}}\right]^{\alpha(2n-1)a/2}I_{1}^{\alpha(1-a)/2}
\,d\tau\nonumber\\
&\leq& c_{n,m} \left[\int_{0}^{t} I_{n}^{\frac{1}{2n-1}}\,d\tau\right]^{\alpha(2n-1)a/2}
\left[\int_{0}^{t} I_{1}^{\frac{\alpha(1-a)}{2 - (2n-1)a\alpha}}\right]^{1- \alpha(2n-1)a/2}d\tau\,.
\eeq
The only way we know to bound the right hand side is to choose $\alpha$ such that the exponent of $I_1$ is unity, namely 
\bel{ln15}
\frac{\alpha(1-a)}{2 - (2n-1)a\alpha} = 1\qquad\Rightarrow\qquad \alpha = \frac{2m}{4m-3}\,.
\ee
Thus $\alpha$ is a function of $m$ only and is uniform in $n$. Thus we label it as $\alpha_{m}$.
\hfill $\Box$


\section{\large The triangular H\"older inequalities for $\Omega_{m}$ and $D_{m}$}\label{appA}

Consider the definition of $P_{m}$ 
\bel{app1}
P_{m}^{2m} = \I |\brot|^{2m}\,dV \equiv \I|\brot|^{2\alpha}|\brot|^{2\beta}dV
\ee
where $\alpha + \beta = m$. Then, for $m > 1$ and $1 \leq p \leq m-1$ and $q > 0$, we have 
\bel{app2}
P_{m}^{2m} \leq  \left(\I |\brot|^{2(m-p)}\,dV\right)^{\frac{\alpha}{m-p}}
\left(\I|\brot|^{2(m+q)}dV\right)^{\frac{\beta}{m+q}}
\ee
where $\frac{\alpha}{m-p} + \frac{\beta}{m+q} = 1$. Solving for $\alpha,~\beta$ gives 
\bel{app3}
\alpha = \frac{q(m-p)}{p+q}\qquad\mbox{and}\qquad \beta = \frac{p(m+q)}{p+q}\,,
\ee
thereby giving
\bel{app4}
P_{m}^{m(p+q)} \leq P_{m-p}^{q(m-p)}P_{m+q}^{p(m+q)}\,.
\ee
Now choose $q=1$ and $p = m-1$ to obtain
\bel{app5a}
P_{m}^{m^2} \leq P_{1}P_{m+1}^{m^{2}-1}\,.
\ee 
Likewise, by the same argument, 
\bel{app5b}
Q_{m}^{m^2} \leq Q_{1}Q_{m+1}^{m^{2}-1}\,,
\ee 
and so using the fact that $\Omega_{m} = P_{m}+Q_{m}$ we also have
\bel{app5c}
\Omega_{m}^{m^2} \leq 2^{m^2}\Omega_{1}\Omega_{m+1}^{m^{2}-1}\,.
\ee 
With the definitions $\rho_{m} = \twothirds m(4m+1)$ and $\eta_{m} = \frac{2m}{3(m-1)}$ defined in (\ref{rhodef})  
we can transform into the $D_{m}$-format to obtain
\bel{app6}
\left(\frac{D_{m}}{D_{1}}\right) \leq 2^{2m^{2}}\left(\frac{D_{m+1}}{D_{m}}\right)^{(4m+1)(m-1)}\,,
\ee
or
\bel{app7}
\left(\frac{D_{m}}{D_{1}}\right)^{\eta_{m}} \leq 2^{2m^{2}\eta_{m}}\left(\frac{D_{m+1}}{D_{m}}\right)^{\rho_{m}}\,.
\ee



\section{\large The proof of Lemma \ref{Omlem}}\label{appB}

In the following $c_{m},~c_{1,m,},~c_{2,m}$ and $c_{3,m}$ are constants with respect to space and time but dependent only 
upon $m$. Consider 
\bel{Jmdef}
J_{m} = \I |\brot|^{2m}dV = P_{m}^{2m}\,,
\ee
such that
\bel{Hm1}
\frac{1}{2m}\dot{J}_{m} = \I |\brot|^{2(m-1)}\brot\cdot
\left\{Re^{-1}\Delta\brot + \brot\cdot\nabla\bu - Fr^{-1}\nabla\rho\times\bk\right\}\,dV\,.
\ee 
Bounds on the three constituent parts of (\ref{Hm1}) are dealt with in turn, 
culminating in a differential inequality for $J_{m}$. 
\par\medskip\noindent
\textit{a)~The Laplacian term\,:} Let $\phi = \omega_{rot}^{2} = \brot\cdot\brot$. Then
\beq{s1a}
\I |\brot|^{2(m-1)}\brot\cdot\Delta\brot\,dV 
&=&\I \phi^{m-1}\left\{\Delta\left(\shalf\phi\right) - |\nabla\bom|^{2}\right\}\,dV\nonumber\\
&\leq& \I \phi^{m-1}\Delta\left(\shalf\phi\right)\,dV\,.
\eeq
Using the fact that $\Delta(\phi^{m}) = m\left\{(m-1)\phi^{m-2}|\nabla\phi|^{2} + 
\phi^{m-1}\Delta\phi\right\}$ we obtain
\beq{s1ex1}
\I |\brot|^{2(m-1)}\brot\cdot\Delta\brot\,dV 
&\leq& -\shalf(m-1)\I\phi^{m-2}|\nabla\phi|^{2}\,dV + 
\frac{1}{2m}\I\Delta(\phi^{m})\,dV\nonumber\\
&=& - \frac{2(m-1)}{m^2}\I|\nabla(\omega_{rot}^{m})|^{2}\,dV\,,
\eeq
having used the Divergence Theorem. Thus we have 
\beq{s1ex2}
\I |\brot|^{2(m-1)}\brot\cdot\Delta\brot\,dV \leq \left\{
\begin{array}{cl}
-\I|\nabla\brot|^{2}]\,dV & ~~~~~m=1\,,\\
-\frac{2}{\tilde{c}_{1,m}}\I|\nabla A_{m}|^{2}\,dV & ~~~~~m\geq 2\,.
\end{array}
\right.
\eeq
where $A_{m}= \omega_{rot}^{m}$ and $\tilde{c}_{1,m} = m^{2}/(m-1)$ with equality at $m=1$. 
The negativity of the right hand side of (\ref{s1ex2}) is important and can be dealt with as follows. 
Recalling that $A_{m} = \omega_{rot}^{m}$ allows us to re-write $J_{m+1}$ as
\bel{s3a}
J_{m+1} = \|A_{m}\|_{2(m+1)/m}^{2(m+1)/m}\,.
\ee
A Gagliardo-Nirenberg inequality yields
\bel{s3b}
\|A_{m}\|_{2(m+1)/m} \leq c_{m}\,\|\nabla A_{m}\|_{2}^{3/2(m+1)}
\|A_{m}\|_{2}^{(2m-1)/2(m+1)} + \|A_{m}\|_{2}
\ee
which means that 
\bel{s3c}
J_{m+1} \leq c_{m}\left(\I|\nabla(\omega_{rot}^{m})|^{2}\,dV\right)^{3/2m}J_{m}^{(2m-1)/2m} + J_{m}^{(m+1)/m}\,.
\ee
With the definition of $\beta_{m}$ given in (\ref{betamdef}) we obtain
\beq{Jm1a}
P_{m+1} = J_{m+1}^{1/2(m+1)} 
&\leq& c_{m}\left(\I|\nabla(\omega_{rot}^{m})|^{2}\,dV\right)^{1/\beta_{m}}P_{m}^{(2m-1)/2(m+1)} + P_{m}
\eeq
which converts to
\bel{Jm1b}
c_{m}^{\beta_{m}}2^{-(\beta_{m}-1)}
\I|\nabla(\omega_{rot}^{m})|^{2}\,dV 
\geq \left(\frac{P_{m+1}}{P_{m}}\right)^{\beta_{m}}P_{m}^{2m} - P_{m}^{2m}\,.
\ee
\textit{b)~The nonlinear term in (\ref{Hm1})\,:} After a H\"older inequality, the second term in (\ref{Hm1}) becomes
\beq{s2a}
\I |\brot|^{2m}|\nabla\bu|\,dV
&\leq& \left(\I |\nabla\bu|^{2(m+1)}dV\right)^{\frac{1}{2(m+1)}}
\left(\I |\brot|^{2(m+1)}dV\right)^{\frac{m}{2(m+1)}}\left(\I |\brot|^{2m}dV\right)^{1/2}\non\\
&\leq& c_{m}\left(\I |\brot|^{2(m+1)}dV\right)^{1/2}\left(\I |\brot|^{2m}dV\right)^{1/2}\non\\
&+& Ro^{-1}\left(\I |\brot|^{2(m+1)}dV\right)^{\frac{m}{2(m+1)}}\left(\I |\brot|^{2m}dV\right)^{1/2}\non\\
&\leq& c_{m}\left(J_{m+1}^{1/2}\,J_{m}^{1/2} + Ro^{-1} J_{m+1}^{\frac{m}{2(m+1)}}J_{m}^{1/2}\right)\non\\
&=& c_{m}\left(P_{m+1}^{m+1}P_{m}^{m} + Ro^{-1}P_{m+1}^{m}P_{m}^{m} \right)
\eeq
where the inequality $\|\nabla\bu\|_{p} \leq c_{p} \|\bom\|_{p}$ for $p \in (1,\,\infty)$ has been 
used, which is based on a Riesz transform and necessarily excludes the case $m=\infty$. It also needs 
to be transformed to $\brot$ using $\brot = \bom + \bk Ro^{-1}$.
\par\medskip\noindent
\textit{c)~The buoyancy term in (\ref{Hm1})\,:}  
\beq{f2}
\left|-\I|\brot|^{2(m-1)}\brot\cdot (Fr^{-1}\nabla\rho\times\bk)\,dV\right| 
&\leq& Fr^{-1}\|\brot\|_{2m}^{2m-1}\|\nabla\rho\|_{2m}\non\\
&=& Fr^{-1}P_{m}^{2m-1}Q_{m}\,.
\eeq
\textit{d)~A grouping of all the terms for $\dot{P}_{m}$\,:}  
Together 
with (\ref{s1a}), this makes (\ref{Hm1}) into 
\bel{Hm2}
\frac{1}{2m}\dot{J}_{m} \leq -\frac{Re^{-1}}{\tilde{c}_{1,m}}\I|\nabla(\omega_{rot}^{m})|^{2}\,dV 
+ c_{m}P_{m+1}^{m+1}P_{m}^{m} + c_{m}Ro^{-1}P_{m+1}^{m}P_{m}^{m} + Fr^{-1}P_{m}^{2m-1}Q_{m}\,.
\ee
Converting the $J_{m}$ into $P_{m}$ gives
\beq{Pm1a}
\dot{P}_{m} &\leq& P_{m}\left\{-\frac{Re^{-1}}{c_{1,m}}
\left(\frac{P_{m+1}}{P_{m}}\right)^{\beta_{m}} + c_{2,m}\left(\frac{P_{m+1}}{P_{m}}\right)^{m+1}P_{m} 
+ c_{2,m}Ro^{-1}\left(\frac{P_{m+1}}{P_{m}}\right)^{m}\right\}\non\\
&+& c_{3,m}\left(Re^{-1}P_{m} + Fr^{-1}Q_{m}\right)
\eeq
\textit{e)~An estimate for $\dot{Q}_{m}$\,:} 
Likewise, the equivalent estimate for $\dot{Q}_{m}$ from (\ref{nse6}) is
\beq{Qm1}
\dot{Q}_{m} &\leq& Q_{m}\left\{-\frac{Re^{-1}}{c_{1,m}}\left(\frac{Q_{m+1}}{Q_{m}}\right)^{\beta_{m}} 
+ c_{2,m}\left(\frac{Q_{m+1}}{Q_{m}}\right)^{m}P_{m+1} + c_{2,m}Ro^{-1}\left(\frac{Q_{m+1}}{Q_{m}}\right)^{m}\right\}\non\\
&+& c_{3,m}\left(Re^{-1}Q_{m} + Fr^{-1}P_{m}\right)
\eeq
Noting that $3/4m + 1/2\alpha_{m} = 1$, we find
\beq{Pm1b}
c_{2,m}\left(\frac{P_{m+1}}{P_{m}}\right)^{m+1}P_{m} &=& 
\left\{\frac{Re^{-1}}{c_{1,m}}\left(\frac{P_{m+1}}{P_{m}}\right)^{\beta_{m}}\right\}^{3/4m}
\left\{c_{m} Re^{6\alpha_{m}/4m}P_{m}^{2\alpha_{m}}\right\}^{1/2\alpha_{m}}\non\\
&\leq& \frac{3}{4m}\frac{Re^{-1}}{c_{1,m}}\left(\frac{P_{m+1}}{P_{m}}\right)^{\beta_{m}} +
\frac{c_{m}}{2\alpha_{m}} Re^{6\alpha_{m}/4m}P_{m}^{2\alpha_{m}}
\eeq
\beq{Pm1c}
c_{2,m}Ro^{-1}\left(\frac{P_{m+1}}{P_{m}}\right)^{m} &=& 
\left\{\frac{Re^{-1}}{c_{1,m}}\left(\frac{P_{m+1}}{P_{m}}\right)^{\beta_{m}}\right\}^{3/4(m+1)}
\left\{c_{1,m}Ro^{-2\alpha_{m+1}}Re^{\frac{3}{4m+1}}\right\}^{1/2\alpha_{m+1}}\non\\
&\leq& \frac{3}{4(m+1)}\frac{Re^{-1}}{c_{1,m}}\left(\frac{P_{m+1}}{P_{m}}\right)^{\beta_{m}} +
\frac{c_{1,m}}{2\alpha_{m+1}}Ro^{-2\alpha_{m+1}}Re^{\frac{3}{4m+1}}
\eeq
and
\beq{Qm2}
c_{2,m}\left(\frac{Q_{m+1}}{Q_{m}}\right)^{m}P_{m+1}Q_{m} 
&=& \left\{\frac{Re^{-1}}{c_{1,m}}Q_{m}\left(\frac{Q_{m+1}}{Q_{m}}\right)^{\beta_{m}}\right\}^{m/\beta_{m}}
\left\{\frac{Re^{-1}}{c_{1,m}}P_{m}\left(\frac{P_{m+1}}{P_{m}}\right)^{\beta_{m}}\right\}^{1/\beta_{m}}\non\\
&\times& 
\left\{c_{m}Re^{6\alpha_{m}/4m}P_{m}^{2\alpha_{m}(1-\beta_{m}^{-1})}Q_{m}^{2\alpha_{m}(1-m\beta_{m}^{-1})}\right\}^{1/2\alpha_{m}}\non\\
&\leq& \frac{m}{\beta_{m}}\frac{Re^{-1}}{c_{1,m}}Q_{m}\left(\frac{Q_{m+1}}{Q_{m}}\right)^{\beta_{m}} + 
\frac{1}{\beta_{m}}\frac{Re^{-1}}{c_{1,m}}P_{m}\left(\frac{P_{m+1}}{P_{m}}\right)^{\beta_{m}} \non\\
&+& \frac{c_{m}}{2\alpha_{m}}Re^{6\alpha_{m}/4m}P_{m}^{2\alpha_{m}(1-\beta_{m}^{-1})}Q_{m}^{2\alpha_{m}(1-m\beta_{m}^{-1})}
\eeq
Also noting that 
\bel{Qm3}
2\alpha_{m}(1-\beta_{m}^{-1}) + 2\alpha_{m}(1-m\beta_{m}^{-1}) = 2\alpha_{m} + 1
\ee
then the application of (\ref{Pm1b}), (\ref{Pm1c}) and (\ref{Qm2}) allows us to write a combination of 
(\ref{Pm1a}) and (\ref{Qm1}) in terms of $\Omega_{m}$ for $m > 1$
\beq{Om1}
\dot{\Omega}_{m} &\leq& -\frac{Re^{-1}}{c_{1,m}}\left(1 - 3/2m -3/4(m+1) - 1/\beta_{m}\right)P_{m}
\left(\frac{P_{m+1}}{P_{m}}\right)^{\beta_{m}}\non\\ 
&-& \frac{Pr^{-1}Re^{-1}}{c_{1,m}}\left(1-m\beta_{m}^{-1} - 3/4(m+1)\right)Q_{m}\left(\frac{Q_{m+1}}{Q_{m}}\right)^{\beta_{m}}\non\\ 
&+& c_{2,m}Re^{6\alpha_{m}/4m}\Omega_{m}^{1+2\alpha_{m}} + c_{3,m}\left(Re^{-1}+ Fr^{-1} 
+ Ro^{-2\alpha_{m+1}}Re^{\frac{3}{4m+1}}\right)\Omega_{m}\,.
\eeq
Now
\bel{Om1a}
1 - 3/2m -3/4(m+1) - 1/\beta_{m} = 1 - 3/2m\qquad\qquad 1-m\beta_{m}^{-1} - 3/4(m+1) = 1 - 3/2(m+1)\,.
\ee
Moreover, using the fact that $\beta_{m} = \fourthirds m(m+1)> 1$ for $m > 1/2$, we find 
\bel{Om2}
\min(Pr^{-1},1)\Omega_{m+1}^{\beta_{m}} \leq 2^{\beta_{m} -1}\left(P_{m+1}^{\beta_{m}} + Pr^{-1} Q_{m+1}^{\beta_{m}}\right)
\ee
together with the two facts that $6\alpha_{m}/4m = 2\alpha_{m} - 1$ and $3/2m > m/\beta_{m}$, (\ref{Om1}) 
becomes
\beq{Om3b}
\dot{\Omega}_{m} &\leq& -\frac{Re^{-1}}{c_{1,m}}2^{\beta_{m} -1}\min(Pr^{-1},\,1)\left(1 - \frac{3}{2(m+1)}\right)\Omega_{m}
\left(\frac{\Omega_{m+1}}{\Omega_{m}}\right)^{\beta_{m}}\non\\ 
&+& c_{2,m}Re^{6\alpha_{m}/4m}\Omega_{m}^{1+2\alpha_{m}} + c_{3,m}\left(Re^{-1}+ Fr^{-1} 
+ Ro^{-2\alpha_{m+1}}Re^{\frac{3}{4m+1}}\right)\Omega_{m}\,.
\eeq
Given that $1 > 3/2(m+1)$ for $m > 1$, and after absorbing the multiplicative constants into $c_{1,m}$ 
we have the result of the Lemma. $\blacksquare$


\end{document}